# CONFIGURATION IN ERP SAAS MULTI-TENANCY

Djamal Ziani[1]

[1]King Saud University, College of Computer and Information Systems Sciences,
Information Systems Department, PO Box 51178 Riyadh 11543 Saudi Arabia
dziani@ksu.edu.sa

## ABSTRACT

*Software as a Service (SaaS) becomes in this decade the focus of many enterprises and research. SaaS provides software application as Web based delivery to server many customers. This sharing of infrastructure and application provided by Saas has a great benefit to customers, since it reduces costs, minimizes risks, improves their competitive positioning, as well as seeks out innovative. SaaS application is generally developed with standardized software functionalities to serve as many customers as possible. However many customers ask to change the standardized provided functions according to their specific business needs, and this can be achieve through the configuration and customization provided by the SaaS vendor. Allowing many customers to change software configurations without impacting others customers and with preserving security and efficiency of the provided services, becomes a big challenge to SaaS vendors, who are oblige to design new strategies and architectures. Multi-tenancy (MT) architectures allow multiple customers to be consolidated into the same operational system without changing anything in the vendor source code. In this paper, we will present how the configuration can be done on an ERP web application in a Multi-Tenancy SaaS environment.*

## 1. INTRODUCTION

Cloud Computing has brought the opportunity to small and medium-size companies to access high processing capabilities so far reserved to big corporations [1]. New applications are developed and offered by software vendors through Internet as a service, thus the customers can access them through web browsers anytime and anywhere. This new form of software distribution is called Software as a Service (SaaS) [2]. The association of commercial software products with SaaS will become the dominant trend for future software development and deployment.

SaaS offers software products, such as Enterprise Resource Planning (ERP) systems to multiple users as services available through the Internet. Users could pay only a subscription fee to SaaS vendors to use these software systems.

Even that, the vendors adapt the best practices modelled in the ERP system, in order that the provided functions can be used by a huge number of customers, however many customers ask to tailor some ERP functions to their business needs. Thus, to answer to the customer requests, and narrow the gap between company-specific business processes and system-embedded best practices, the ERP vendors provide to their customers tools to do their own customizing and configuration, by using Multi-tenancy Architectures (MTA) [3]. Multi-tenancy is an architectural pattern in which a single instance of the software is run on the service provider's infrastructure, and multiple tenants access the same instance. On the other hand, in single-tenant environment every tenant has his own customized application instance [4].

Traditional multi-tenant applications are shared among tenants with common functional needs. This solution satisfy partially some customers, since most of their specific needs can be covered. However, to satisfy all customers and provide a flexible solution, the ERP vendors should provide a Configuration in SaaS as services, and thus use a new Multi-tenant architecture.

## 2. CLOUD COMPUTING

First lets define what is Cloud Computer, in the literature we can find many definitions, so in this paper we will Berkeley's definition. Berkeley defines Cloud Computing as the sum of Utility Computing and Software as a Service [1]. Utility Computing represents the use of computer resources on demand and it enables a distribution formula for software vendors called Software as a Service [5]. Applications are installed in the software Vendor servers and accessed using Internet. Companies who use the software are not owners of the software, but consumers of web applications.

### 2.1. Computing Architecture

There are three cloud computing layers [6]: Infrastructure as a Service (IaaS), Platform as a Service (PaaS), and Software as a Service (SaaS), as shown in figure 1.

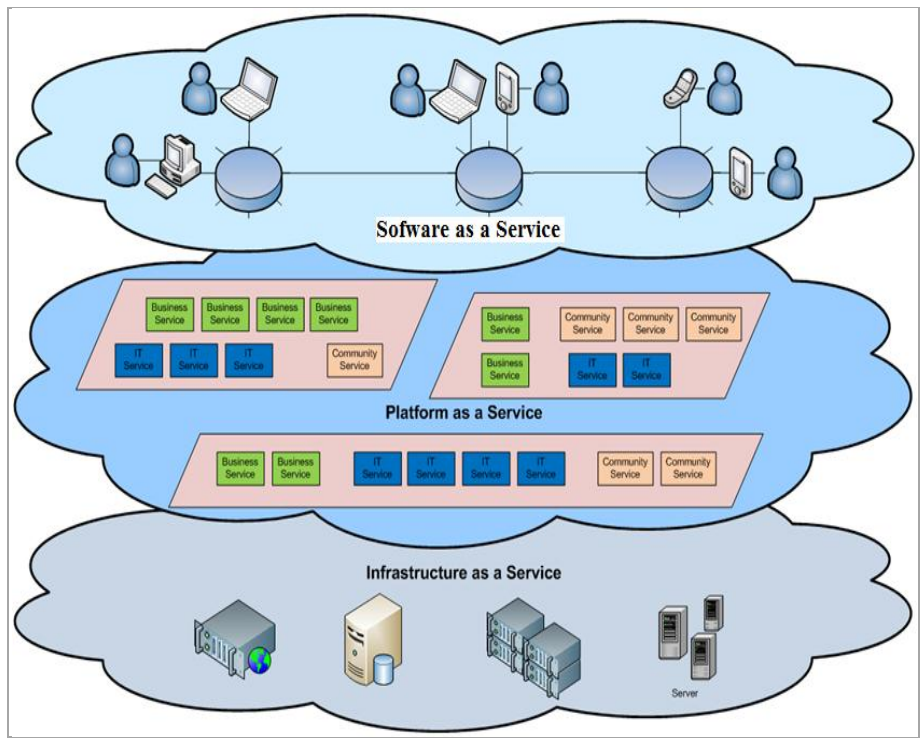

Figure 1. Cloud Computing Layers

- Infrastructure as a Service (IaaS)

Infrastructure as a Service is a provision model in which an organization outsources the equipment used to support operations, including storage, hardware, servers and networking components. The service provider owns the equipment and is responsible for housing, running and maintaining it. The client typically pays on a per-use basis.

- Platform as a Service (PaaS)

Provides a computing platform and a solution stack as a service. The consumer creates the software using tools and/or libraries from the provider. The consumer also controls software deployment and configuration settings.

- Software as a Service (SaaS)

Software as a Service (SaaS) is a software distribution model in which applications are hosted by a vendor or service provider and made available to customers over a network, typically the Internet. It Supports technologies such as Web services and service-oriented architecture (SOA).

## 2.2. Cloud Deployment Models

The most common types of cloud computing deployment models, according to the National Institute of Standards of Technology [7] (see figure 2), are:

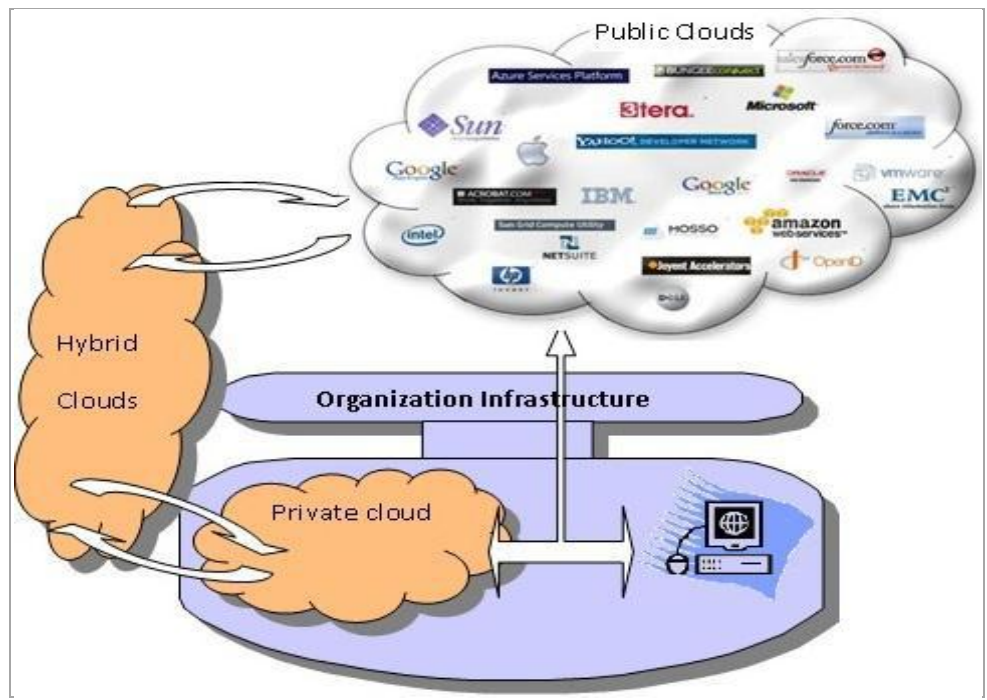

Figure 2.  Cloud Computing Models

- Private cloud

The cloud infrastructure is operated solely for an individual organization and managed by the organization or a third party; it can exist on or off the organization's premises.

- Community cloud

The cloud infrastructure is shared by several organizations and supports a specific community that has common interests (e.g., mission, industry collaboration, or compliance requirements). It might be managed by the community organizations or a third party and could exist on or off the premises.

- Public cloud

The cloud infrastructure is available to the general public or a large industry group and is owned by an organization selling cloud services.

- Hybrid cloud

The cloud infrastructure is composed of two or more clouds (private, community, or public) that remain unique entities but are bound together by standardized or proprietary technology that enables data and application portability.

## 3. ERP SYSTEM

Enterprise Resource Planning (ERP) System coordinates all resources and integrates information across an entire enterprise to ease the flow of information among the different processes within an enterprise, and facilitates the communication and information sharing across organizational units. Implementing an ERP system in enterprise achieves many tangible and intangible benefits; it enables an enterprise to function as a single unit rather than as several separate units and work processes [9]. A well-constructed and implemented ERP can increase productivity and provide information reliability, data redundancy prevention, cost lowering, and scalability and a global outreach improvement. ERP is one of the largest IT investment that adopted by organizations since more than a decade.

To cover all business activities of a company, an ERP system contains various modules which are tightly integrated and collaborate together to perform the business scenarios of a company. Figure 3 shows the most important modules.

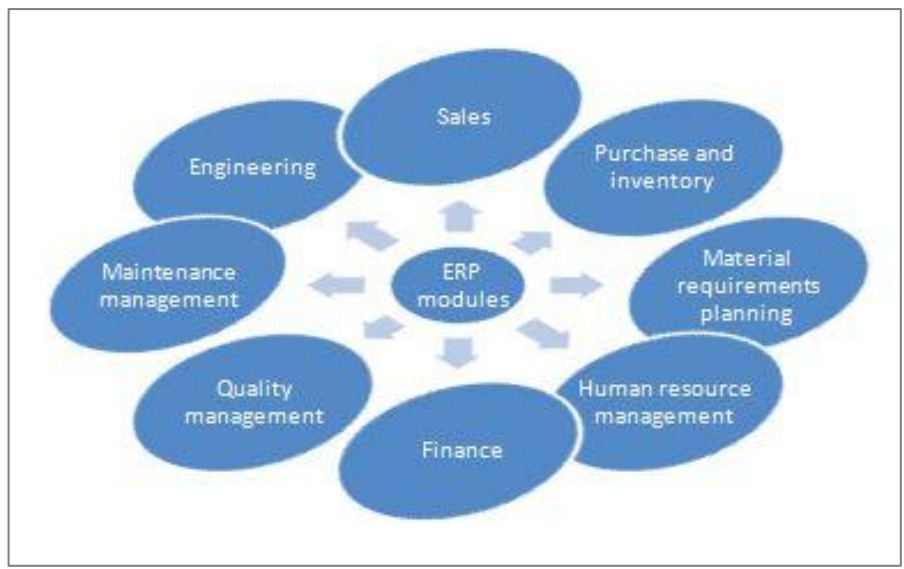

Figure 3. ERP modules

The ERP systems are based on good strong architectures that allow them to be flexible, efficient and secure. Figure 4 shows an architecture of Web SAP CRM application. Since SAP is the biggest ERP software company in the world, we will base on this architecture to explain how Multi-Tenant configuration in SaaS application can be done.

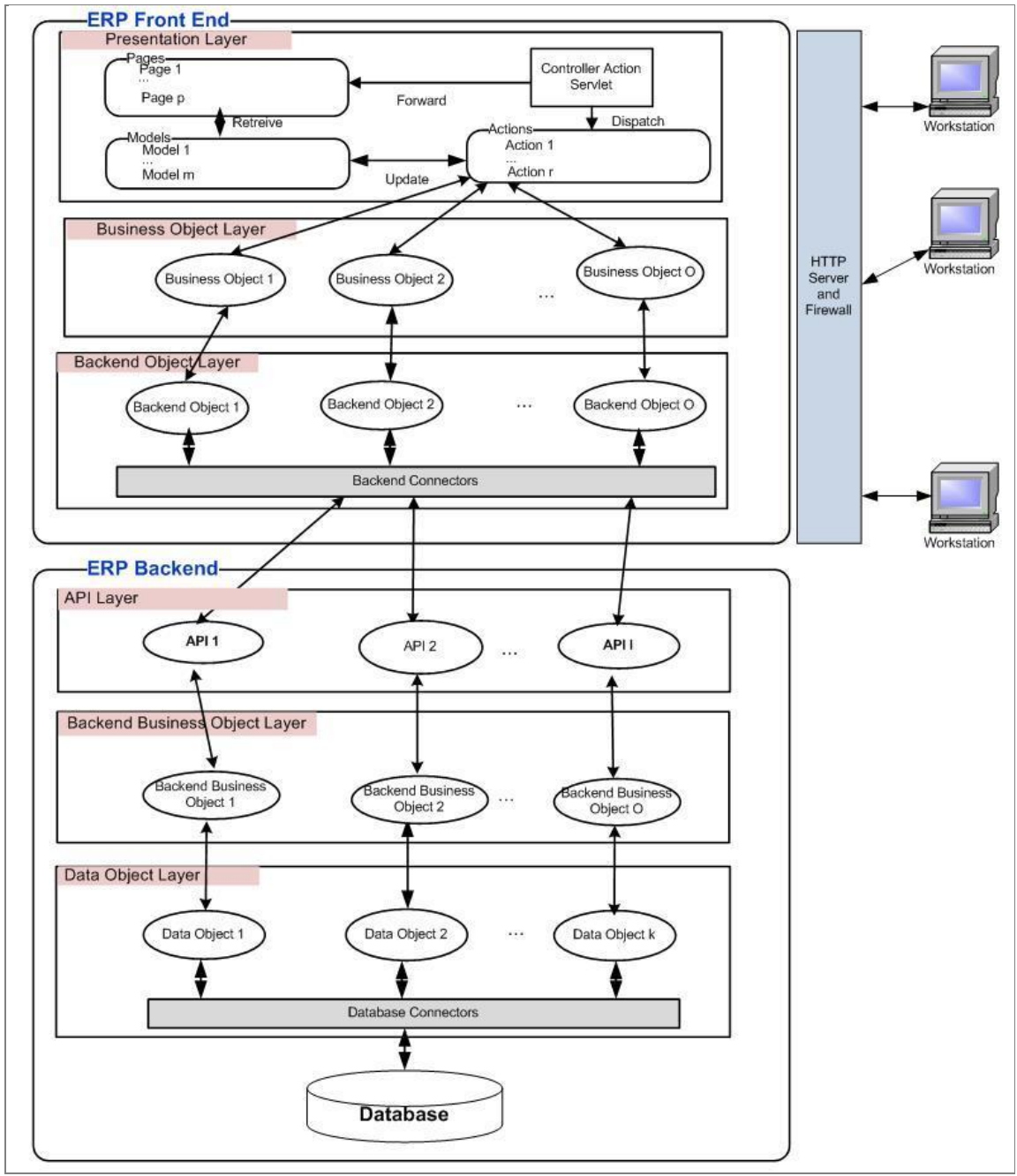

Figure 4. Web SAP CRM architecture

- ERP Frontend: It is organized using three layers.

  o Presentation layer: the objective of this layer is to interact directly with the end users. This layer is architecture using the MVC (Model View Control) paradigm. The view in MVC is represented here by the web pages, the model represents a container object which is used to save and populated the data of a view (here the web page), and the controller is represented by a set of actions, which are object classes that execute the requests performed from the web pages.

  o Business Object layer: In order to use any backend system, the business logic is dividing into a backend-independent (Business Objects) and a dependent part (Backend

Objects). The Business Objects (business logic objects or BOs) use a generic set of interfaces to communicate with the backend, therefore their purpose is to represent the backend-independent part of the business logic.

- o Backend Object layer: Since all Business Objects have to be independent from the backend system, they have to access the Backend Object through a unified interface. The implementation of this interface is backend-dependent, but Business Objects are not aware if this. Since the implementation of a Backend Object can change. The backend objects have as mission to call backend API using backend connector objects.

- ERP Backend: It is organized using three layers.

  - o API layer: APIs (Application Programming Interfaces are specifications for routines, data structures, object classes, and variables to accomplish the communication between the ERP frontend layer and the ERP backend. They are called from frontend backend objects to execute a specific task in the backend.

  - o Backend Business Object layer: The Backend business object represents functional areas of the enterprise, for example: Sales, Procurement, Opportunity, Account, and Service. These objects contain the core of the enterprise business as a set of routines or classes.

  - o Data Object layer: The objective of the objects of this layer is to get or set information from or to database and send it to a comprehensive form to the caller of backend business object layer.

## 4. RELATED WORK

In the literature there is a lack of clear identification of the difference between configuration and customization in SaaS systems. However, Sun et al. provided a simple description about the difference between configuration and customization in SaaS from the vendor perspective [9]. They also presented a methodology framework to guide SaaS vendor to plan and execute configuration and customization strategies.

Configuration is one of the key success of any SaaS software. It aims to give each tenant (client) the ability to modify its system according to their needs in a pre-defined options and variations. The configuration approach doesn't involve source code changes.

Customization requires source code modification. It is one of the main problems that organizations complained about. According to Sun et al., customization is more complex and costly because it involves changing in the source code of SaaS software. That will require higher skills consultants and involve much longer lifecycle to develop, debug, test and deploy the code [10].

Very few research papers have student the configuration in ERP SaaS Multi-Tenancy environment, we can cite:

- Nitu proposed SaaS application architecture to support configuration by the tenant's designer. He defined two main types of user: tenant's designer and tenant's end user; each of them should have its own application module. To configure the system a set of predefined templates are used. By using the UI designer, the tenant's designer can use these templates in the application edit mode, while the end user uses the configured data in the play mode of the application [11]. Unfortunately, this paper didn't discuss how to do the configuration in a multi-tenancy scenario.

- Mietzner et al. proposed in their paper the use of variability descriptors and multi-tenancy patterns to extend the service component architecture (SCA) [12]. In their approach they the application in two parts: object shared by all tenants (common and non configurable objects), and tenant specific data. The configuration and customization in there package format is limited to the properties that delivered within the SCA specification, and they did not explain how to make changes in the complex configuration, such as ERP system configuration.

- Arya et al. explained int heir paper the two role of users of SaaS application, tenant designer which is the responsible of system configuration and the tenant end user, which will use the application. They described the category of objects that can be configured in SaaS application (UI, workflow, business processes flow), and they suggested a metadata customization module for each of these objects category [13]. This paper did not provide enough technical details of the configuration process, especially regarding the workflow and business processes flow.

- Bezemer et al. multi-tenant have categorized the object that can be configured in Saas Multi-Tenant environment: layout Style - to allow the use of tenant-specific themes and styles; general configuration allows the specification of tenant-specific configuration, such as personal profile details; file I/O to allow the specification of tenant-specific file paths, which can be used for, e.g., report generation; and workflow to allows the configuration of tenants specific workflows, for instance ERP workflow [14]. They give high level architecture, but the paper did not provides enough detail information on how a configuration can change from a single tenant environment to a multi-tenant environment.

- Kang et al. Proposed conceptual architecture of a SaaS platform that enables executing of configurable and multitenant SaaS application. In their paper they explained what are the different objects that can be configured in an ERP systems: organizational structure where the tenant's manager can change the role sets based on initial roles that SaaS application developer created; user interface is to change look and feel of the UI; data model means the ability to add/delete data object, or to add/delete data fields in existing data object; workflow allows the tenant manager to configure various components of business process in terms of workflow, activity type, and business rules; business logic allows tenant the manager can create a simple business logic using template class [15]. Also they proposed an architecture based on metadata where each tenant can have its own specific metadata. Furthermore, the paper suggests a conceptual architecture for SaaS platform, where each component have been explained.

## 5. CONFIGURATION OF SAAS ERP MULTI-TENANCY

### 5.1. ERP Configuration Objects

The big ERP vendors, such as SAP, Oracle, Sage, etc. offer configuration tools that permit to tailor ERP applications to the needs of companies. A good configuration should convert all the layers of the application. In our paper we will present the different configuration that SAP web application provide to the companies.

#### 5.1.1. Configuration of ERP Frontend Presentation layer

A web page contains several objects that can be configured:

**a. Cascading style sheet (CSS):** CSS is designed to enable the separation of document content from document presentation. Using a CSS a customer can change the default provided UI elements such as the layout, colors, and fonts. To configure the CSS the web page contain the names of CSS used, and the location of CSS file is location in a configuration XML file as follows:

```
<CSSELEMENT>
   <Name>B2C</NAME>
   <LOCATION>"/path1/cssb2c"</LOCATION>
</CSSELEMENT>
<CSSELEMENT>
```

  <Name>B2B</NAME>
  <LOCATION>"/path1/cssb2b"</LOCATION>
 </CSSELEMENT>

**b. Images:** A web page can contain many images, by using a configuration tool a customer can change the images, and get different look and feel. To allow a customer to use its own images, the web page contain only the names of the image; the image path is saved in a configuration XML file, and by using JavaScript function the value of "src" attribute of the html tag <IMG> is loaded. This is image configuration file:

 <IMAGEELEMENT>
  <Name>MyImage</NAME>
  <SRC>"/path1/ image.jpg"</LOCATION>
 </IMAGEELEMENT >

**c. JavaScript files:** JavaScript is a programming language used to make web pages interactive. It allows client-side scripts to interact with the user and control the browser. To have more flexibility, a customer using SAP web application can replace the provide JavaScript files by its owns. Loading dynamically a JavaScript in a web page can be done by using a JavaScript method in the page and by saving the association between a JavaScript name and a JavaScript source in an XML file. This is an example of JavaScript method which load a JavaScript file:

```
function loadjscssfile(filename, filetype){
 if (filetype=="js"){ //if filename is a external JavaScript file
  var fileref=document.createElement('script')
  fileref.setAttribute("type","text/javascript")
  fileref.setAttribute("src", filename)
 }
```

This is an example of XML file:

<SCRIPTELEMENT>
  <Name>MyScript</NAME>
  <SRC>"/path1/MyScript.js"</LOCATION>
</SCRIPTELEMENT>

**d. Property Configuration:** It contains all language-dependant text displayed in the UI, it includes the labels of the fields and the texts displayed in the web pages. A customer can easily use its own label and text by changing the property configuration. To facilitate this type of configuration, each input and text should be loaded from an XML file and not hardcoded in the web page. For instance, by using custom tags textfield and label, as those used by Struts2 for performing localization [6]. The following is an example how the tag is used in web page:

<s:textfield name="Mypage.lastName" label="getText(Mypage.lastName')" ../>

To deal with many languages, the customer can define for language an XML property configuration file, defined as follows:

```xml
<LABELS>
    <LABELELEMENT>
        <Name>Page1.Label1</NAME>
        <VALUE>"My Label 1" </VALUE>
    </LABELELEMENT>

….
    <LABELELEMENT>
        <Name>Page1.Label n</NAME>
        <VALUE>"My Label n" </VALUE>
    </LABELELEMENT>

</LABELS>
<TEXTS>
    <TEXTELEMENT>
        <Name>Page1.Text1</NAME>
        <VALUE>"My Text 1" </VALUE>
    </TEXTELEMENT>

….
    <TEXTELEMENT>
        <Name>Page1. Textm</NAME>
        <VALUE>"My Text m "</VALUE>
    </TEXTELEMENT>

</TEXTS>
```

**e. Section Configuration:** A web page may include many sections, so the customer can hide or collapse any section and also it can show hidden sections.

From SAP CRM 6.0 version, SAP provides to its customers an Easy Enhancement WorkBench UI Configuration. This Tool allows to the customer to enhance the UI by changing the sections used in the presentation layer (see figure 5) [17].

The customer can select, from available assignment blocks, the section that he wants to add to it page. He can also remove any section from a page, change its title or collapse it, see figure 5. The section configuration can be done by using the following XML:

```xml
<BLOCKS>
    <BLOCK>
        <COMPONENT>Component 1</COMPONENT>
        <VIEWNAME>ViewI </VIEWNAME>
        <TITLE>"Block Title 1" </TITLE>
        <DISPLAY>True</DISPLAY>
        <LOADOPTION>Direct</LOADOPTION>
    </BLOCK>

….
    <BLOCK>
        <COMPONENT>Component n</COMPONENT>
        <VIEWNAME>ViewJ </VIEWNAME>
        <TITLE>"Block Title n" </TITLE>
        <DISPLAY>False</DISPLAY>
        <LOADOPTION>Lazy</LOADOPTION>
    </BLOCK>
</BLOCKS>
```

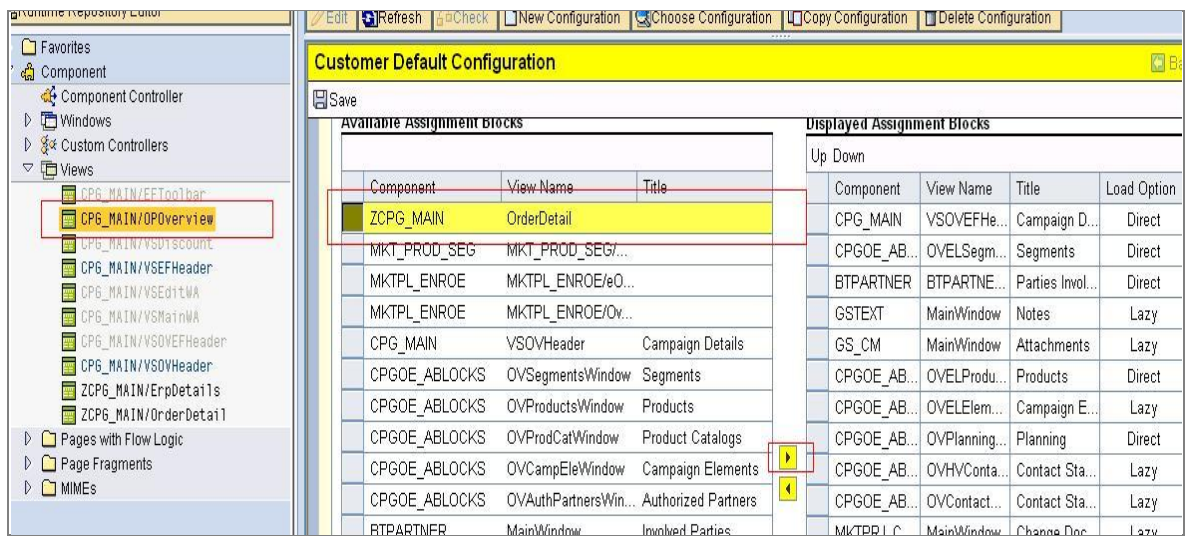

Figure 5. Section Configuration

**f. Field Configuration:** The fields displayed a web page are those provided by default by SAP to all customers. However any customer can hide some fields and also can add in the page some fields from a set of complete page fields provided by the configuration tool.

In the figure 6, we can see, on the left tab of the UI Customer Configuration, all available fields that a customer can add to a page. The customer can also delete a field or change its position in the page. The field configuration can be done by using the following XML:

```
<FIELDS>
    <FIELD>
        <FIELDNAME>Field1</FIELDNAME>
        <DISPLAY>True</DISPLAY>
        <POSITIONFROM>A3 </POSITIONFROM>
        <POSITIONTO>H3</POSITIONTO>
    </FIELD >
….
    <FIELD>
        <FIELDNAME>Fieldn</FIELDNAME>
        <DISPLAY>False</DISPLAY>
        <POSITIONFROM>A11 </POSITIONFROM>
        <POSITIONTO>P11</POSITIONTO>
    </ FIELD>
</ FIELDS>
```

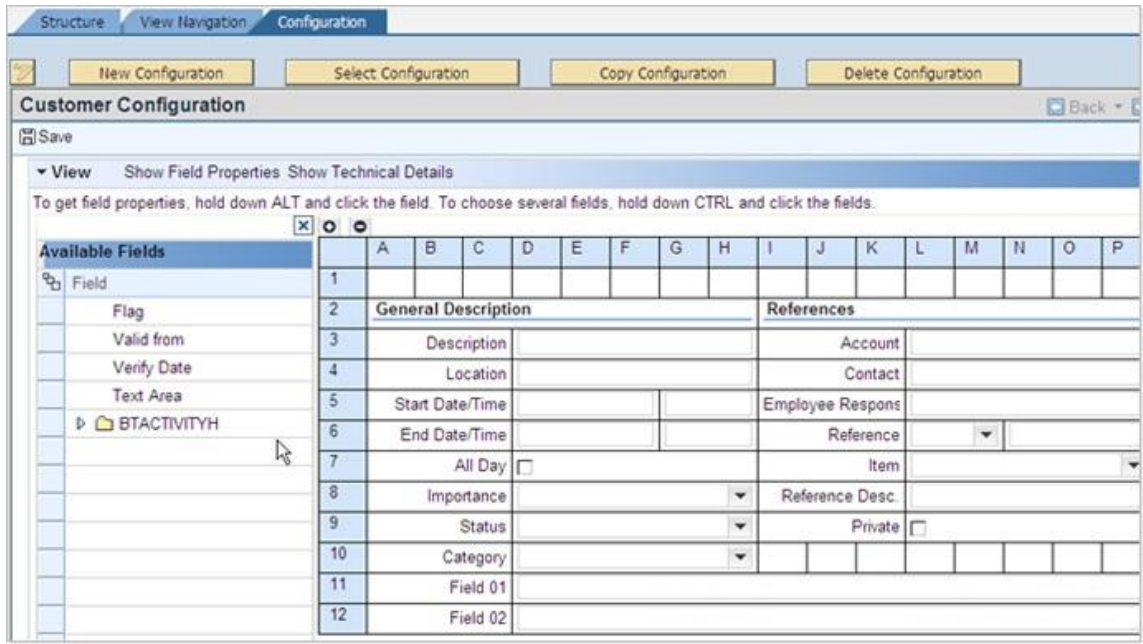

Figure 6. Customer Field Configuration

### 5.1.2. Configuration of ERP Frontend Business Object layer

The only configuration that a customer can do on the Business Object layer of the ERP Frontend is to disable a default provided business objects. Changing or adding a business objects is related to the customizing area. A disabled business object will send an error message to caller action class. The business object configuration can be done using the following XML:

```
<BOS>
    <BO>
        <BONAME>BO1</BONAME>
        <ENABLE>True</ENABLE>
    </BO>
….
    <BO>
        <BONAME>BOn</BONAME>
        <ENABLE>False</ENABLE>
    </BO>
</BOS>
```

### 5.1.3. Configuration of ERP Frontend Backend Object layer

The customer can change the names API called by the backend object classes, and for each API he can also specify which ERP backend to use, and if the connection of API call is statefull (the connection is open during all application live) or stateless (the connection is closed after each call). The backend object configuration can be done using the following XML:

```
<BES>
    <BE>
        <BENAME>BE1</BENAME>
        <API>API1</API>
        <STATE>Full</STATE>
```

```
            <ERPBACKEND>CRM7.0</ERPBACKEND >
        </BE>
    ….
        <BE>
            <BENAME>BEJ</BENAME>
            <API>APIn</API>
            <STATE>Less</ STATE>
            <ERPBACKEND>CRM7.0</ERPBACKEND >
        </BE>
    </BES>
    <CONNECTIONS>
        <CONNECTION>
            <NAME>CRM7</NAME>
            <HOST>CRM7Host</HOST>
            <CLIENT>100</CLIENT>
        </CONNECTION>
    </CONNECTIONS>
```

### 5.1.4. Configuration of ERP Frontend Business Roles

Business Role is a crucial aspect in configuring the Web UI, for instance a Sales employee is able to login as a Sales Professional, and a his manager will login with the business role Sales Manager. The UI objects displayed for an employee and a manager are different, a manager should have more decision report, and the sales employee will have more transactional pages. For instance SAP CRM 7.0 allows linking to a business role to several configuration objects that control the objects accessed by a user by on his role. The configuration objects provides by SAP CRM 7.0 are:

- Navigation Bar Profile: The Navigation Bar profile provides the logical structure to access the internal and external applications within the Web UI.
- Technical Profile: This helps define various technical browser related settings such as disable the use of browser's back button, defining a web page to to load once the user logs off etc.
- Layout Profile: The Layout Profile defines the Navigation Frame of the CRM Web Client. The navigation frame can be used to define the header area, footer area, work area, and navigation bar.
- PFCG Role: This defines the authorization profile for the users, such as what is accessible, editable, etc.

The business role configuration can be organized using the following XML:

```
<BUSINESSROLES>
    <BUSINESSROLE>
        <NAME>SP_ROLE</NAME>
        <DESCRIPTION>"Service Professional Role"</DESCRIPTION>
        <NAVBAR>SRV_NAV_BAR</NAVBAR>
        <TECPROFILE>SRV_TEC_PROFILE</TECPROFILE>
        <LAYPROFILE>SRV_LAY_PROFILE</LAYPROFILE>
        <PFCG>SRV_PFCG</PFCG>
    </BUSINESSROLE>
….
</BUSINESSROLES>
```

### 5.1.5. Configuration of ERP Backend Business Objects

**a. BOL Configuration:** The customers have the flexibility to enable or disable the use of a backend business objects, based on the user business roles. SAP CRM architecture uses the Business object layer (BOL) to save the backend business object data ( for example of sales orders, at runtime of the SAP CRM session), these BOLs can be configured by the customer to restrict the use of a set backend business objects. For instance, a sales representative cannot execute the business objects of financial BOL. The following XML shows how the BOL can be configured:

```xml
<BUSINESSROLES>
    <BUSINESSROLE>
        <NAME>SP_ROLE</NAME>
        <DESCRIPTION>"Sales Professional Role"</DESCRIPTION>
        <BOLS>
            <BOL>
                <NAME>SALES_BOL</NAME>
                <USE>True</USE>
            </BOL>
            <BOL>
                <NAME>FINANCE_BOL</NAME>
                <USE>False</USE>
            </BOL>
            ...
        </BOL>
    </BUSINESSROLE>
….
</BUSINESSROLES>
```

**b. Data Object Configuration:** The customers have the possibility to configure which database to use to perform any request from a backend business object. In general the customer uses only one database, however some customers use their another database for some custom development. To configure this, we use an XML that contain what the customer what to change, the other data objects will use automatically the default database:

```xml
<DOS>
    <DO>
        <NAME>DOMINING</BENAME>
        <DATABASENAME>CRMBI</DATABASENAME>
    </DO>
….
</DOS>
<DATABASES>
    <DATABASE>
        <NAME>CRMDB</NAME>
        <HOST>CRMDBHost</HOST>
        <USE>Default</USE>
    </DATABASE>
    <DATABASE>
        <NAME> CRMBI </NAME>
        <HOST>CRMBIDBHost</HOST>
        <USE>Request</USE>
    </DATABASE>
</DATABASES>
```

**c. Workflow Configuration:** The workflow is basically a business process that consists of a number of sequential tasks performed in a particular order, following a set of rules that is designed to facilitate a particular objective. SAP provides a powerful graphical tool called Workflow Builder to allow the customer to change or build their own workflow. The customer selects the activity types, the tasks, the business object and its method. At the end the workflow should be associated to organization role to specify who can execute this workflow. The figure 7 shows the SAP workflow builder.

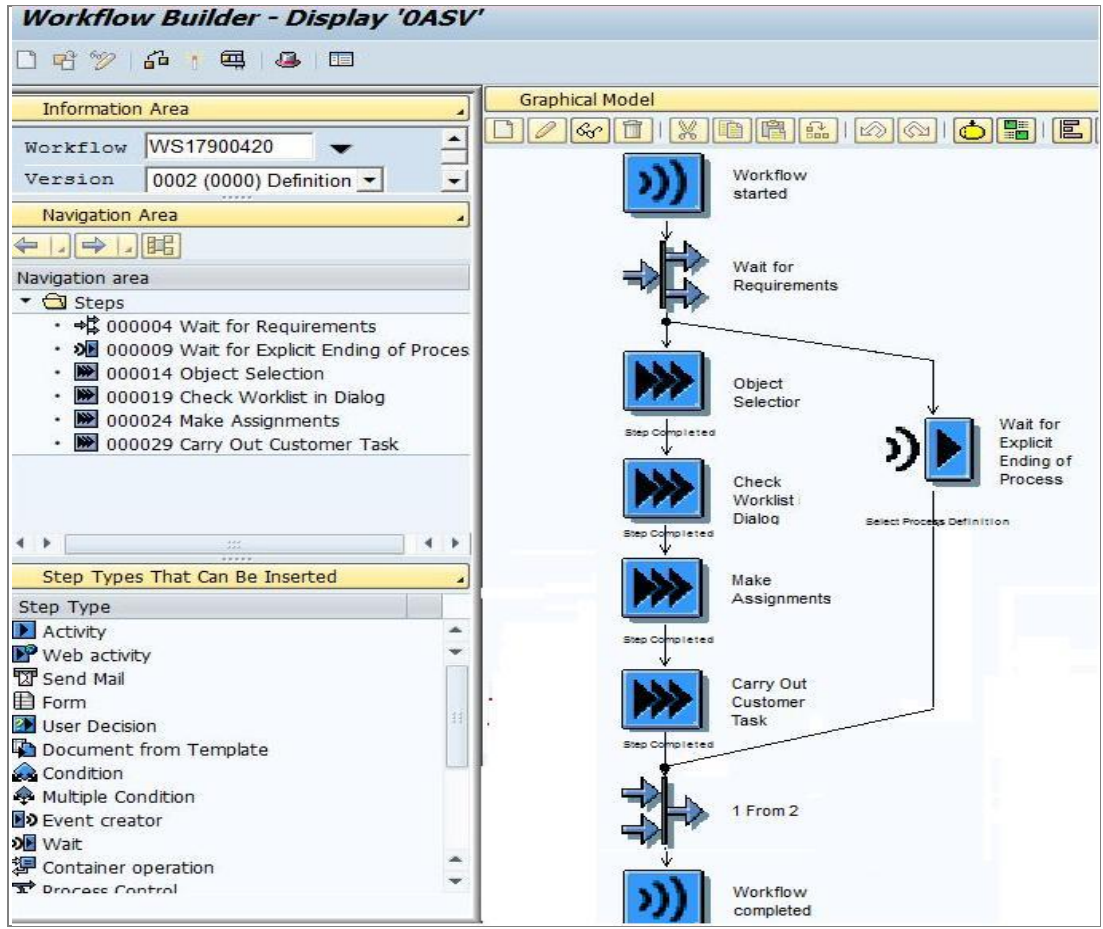

Figure 7. SAP workflow builder

**d. Other Configuration**: SAP provides thousand of other configurations for the customers. In general, these configurations can be of the form <configuration object>=<value>, for example <default Service transaction>=<SO>; or of the form <configuration object>=<set of values>, for example <default Sales business partners> = <{sold-to, ship-to, bill-to, payer}>.

## 5.2. Configuration in SaaS multi-Tenancy

Managing the configuration of many tenants of same application in the same time has two big challenges: the Privacy and security, and the performance.

### 5.2.1. Privacy and Security in Multi-Tenancy SaaS Configuration

The tenant can see only its own configuration, which is saved in a secure directory with a restricted access. Also, the configuration of any tenant will not influence the system of other tenants. To realize this, all configuration files should be read only by using a secure multi-tenant configuration reader. To configure his application the tenant will use a configuration application that allows him to select the category of configuration object to configure (see figure 8). The name and the logo of company tenant is displayed dynamically on the configuration application page. When the tenant client on the "Configure" button the corresponding tenant configuration file will be displayed (a copy of the SAP default configuration file, if it is the first time that the tenant configure this category of objects.

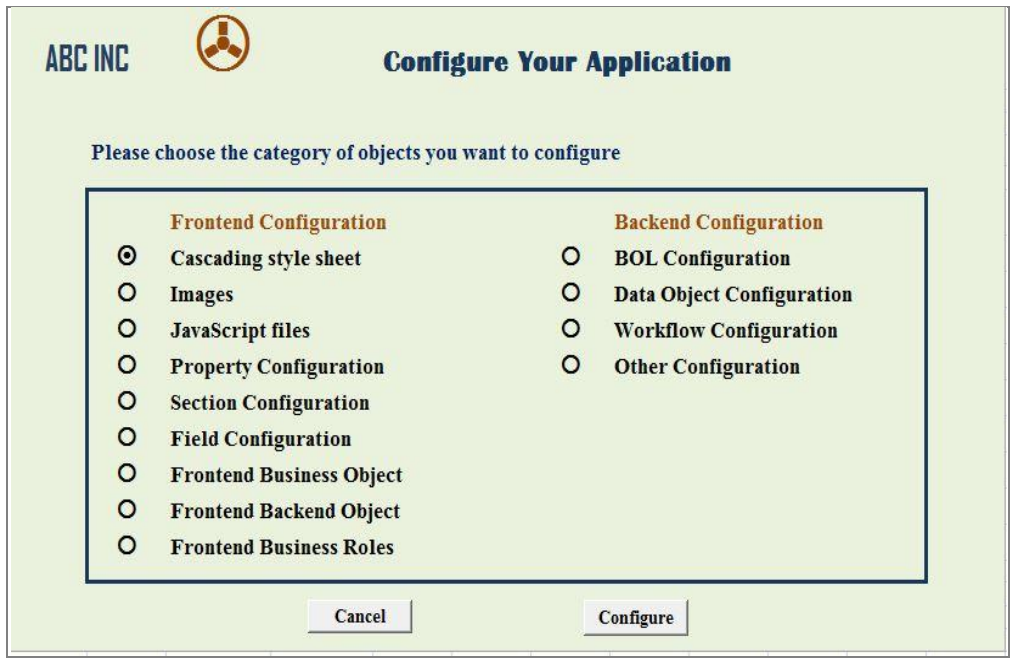

Figure 8.  Prototype of Multi-Tenancy Configuration Application

A central multi-tenant configuration file is used to save to the localization of each tenant configuration. This file includes sections that represent the categories of object configuration, and under each section the file contains the localization of SAP default configuration, and the localization of the tenant configuration file. For more security, this file is not accessible by the tenants, only the ERP cloud provider who can read this file. The following shows the structure of a central multi-tenant configuration file:

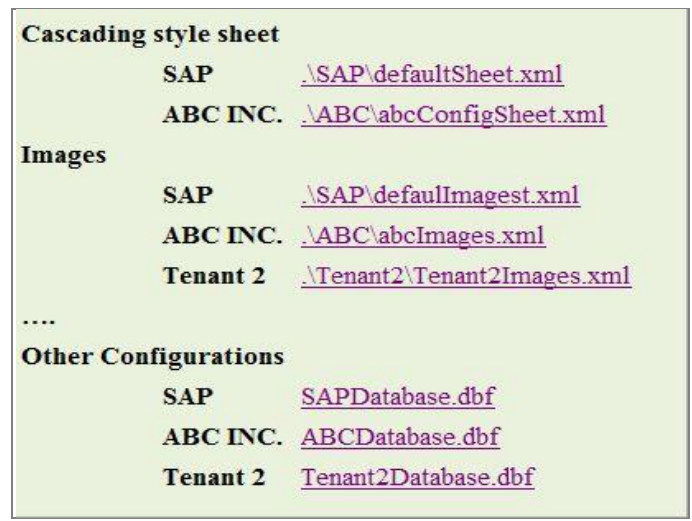

Figure 9.  A central multi-tenant configuration file

Also, we know that SAP uses special table types "Customizing tables" to save most of the backend configuration data. These tables are saved in the same database use to save transactional data. So, to ensure the privacy and the security of the tenant configuration data, we suggest that each tenant will use a separate database to save the transactional and the configuration data.

### 5.2.2. Performance in Multi-Tenancy SaaS Configuration

In multi-Tenancy SaaS environment, without a good architecture and strategy the application performance can be a big issue for cloud providers. To permit a good performance to its clients SAP is using the following techniques:

**a. Tenant configuration isolation:** As seen in the previous subsection, the tenant configurations are separate. The system will first check in the central multi-tenant configuration file where is located the configuration tenant file, and then read the file. It means the system will deal with small configuration files.

**b. Tenant database isolation:** Having different database for each tenant will surely speedup the read of the configuration which are saved in the database side.

c. **Multi layer architecture vs. configuration:** SAP web application is based on a multi layer architecture; thus, the configuration of the application is most of the time organized in a way to change only one layer, this will speed up the load the application changes.

d. **Dynamic integration of the configuration in the development objects:** Most of the configurations are taken into account by using development object variables, getting their values by reading xml configuration files or database configuration tables. These variables can be used in a test to choose the right development peace to execute based on a configuration value, or can be also be embedded in a web page to display or hide dynamically an html code. So this means, based on this technique, the tenant application will be compiled when a change occurs in a configuration.

## 6. CONCLUSIONS

In this paper, we have presented what flexibility and configuration SAP provides to its customers to configure their web applications. We have seen that the configuration can change all application layers.
We presented also how the configuration is processed in a multi tenancy environment. We treated the security and privacy issue by proposing a secure multi-tenancy configuration application which is using a central configuration file and by using the technique of creating a copy of the default SAP category configuration file for a tenant when he needs to change a category of configuration objects.
The performance issue has also been treated by listing all techniques that SAP is using to provide a high performance configuration solution for multi-tenancy SaaS environment. Among these techniques we can cite tenant configuration isolation and tenant database isolation.

## ACKNOWLEDGEMENTS

The authors would like to thank King Saud University, the college of Computer and Information Sciences, and the Research Center for their sponsorship.

**Authors**

Djamal Ziani is Professor Assistant in King Saud University in Computer Sciences and Information Systems College from 2009 until now. Researcher in data management group of CCIS King Saud University. He received Master degree in Computer Sciences from University of Valenciennes France in 1992, and PH.D. in Computer Sciences from University of Paris Dauphine, France in 1996.

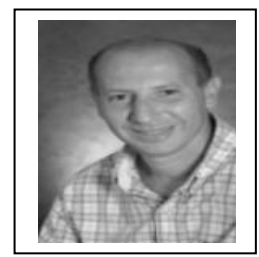